\documentclass[twocolumn,twoside,slac_two]{revtex4}
\usepackage{graphicx}
\usepackage{fancyhdr}
\begin{document}

\title{Sensitivity of KASCADE-Grande data to hadronic interaction models}

\author{D. Kang$^{a \footnote{corresponding author, e-mail: donghwa.kang@kit.edu}}$,
W.D. Apel$^{b}$, J.C. Arteaga-Vel\'azquez$^{a \footnote{now at: Instituto de
    F\'\i sica y Matem\'aticas, Universidad Michoacana, Morelia, Mexico}}$, 
K. Bekk$^{b}$, M. Bertaina$^{c}$, J. Bl\"umer$^{a,b}$, H. Bozdog$^{b}$,\\
I.M. Brancus$^{d}$, P. Buchholz$^{e}$, E. Cantoni$^{c}$, A. Chiavassa$^{c}$,
F. Cossavella$^{a \footnote{now at: Max-Planck-Institut f\"ur Physik, M\"unchen, Germany}}$,
K. Daumiller$^{b}$,\\
V. de Souza$^{a \footnote{now at: Universidade S$\tilde{a}$o Paulo,
Instituto de Fisica de S$\tilde{a}$o Carlos, Brasil}}$, F. di Pierro$^{c}$,
P. Doll$^{b}$, R. Engel$^{b}$, J. Engler$^{b}$, M. Finger$^{a}$,
D. Fuhrmann$^{f}$, P.L. Ghia$^{g}$,\\ 
H.J. Gils$^{b}$, R. Glasstetter$^{f}$, C. Grupen$^{e}$, A. Haungs$^{b}$,
D. Heck$^{b}$, J.R. H\"orandel$^{a \footnote{now at: Dept. of Astrophysics,
Radboud University Nijmegen, The Netherlands}}$, T. Huege$^{b}$,\\
P.G. Isar$^{b \footnote{now at: Institute for Space Sciences, Magurele-Bucharest, Romania}}$, 
K.-H. Kampert$^{f}$, D. Kickelbick$^{e}$,
H.O. Klages$^{b}$, K. Link$^{a}$, P. {\L}uczak$^{h}$, M. Ludwig$^{a}$,\\
H.J. Mathes$^{b}$, H.J. Mayer$^{b}$, M. Melissas$^{a}$, J. Milke$^{b}$,
B. Mitrica$^{d}$, C. Morello$^{g}$, G. Navarra$^{c \footnote{deceased}}$,\\
S. Nehls$^{b}$, J. Oehlschl\"ager$^{b}$, 
S. Ostapchenko$^{b \footnote{now at: University of Trondheim, Norway}}$,
S. Over$^{e}$, N. Palmieria$^{a}$, M. Petcu$^{d}$, T. Pierog$^{b}$,\\ 
H. Rebel$^{b}$, M. Roth$^{b}$, H. Schieler$^{b}$, F.G. Schr\"oder$^{b}$,
O. Sima$^{i}$, G. Toma$^{d}$, G.C. Trinchero$^{g}$,\\ 
H. Ulrich$^{b}$, A. Weindl$^{b}$, J. Wochele$^{b}$, M. Wommer$^{b}$,
J. Zabierowski$^{h}$}

\affiliation{
$^{a}$Institut f\"ur Experimentelle Kernphysik, Karlsruher Institut f\"ur
Technologie - Campus S\"ud, 76021 Karlsruhe,\\ Germany\\ 
$^{b}$Institut f\"ur Kernphysik, Karlsruher Institut f\"ur Technologie - Campus Nord,
76021 Karlsruhe, Germany\\
$^{c}$Diparimento di Fisica Generale dell'Universit\`a, 10125 Torino, Italy\\
$^{d}$National Institute of Physics and Nuclear Engineering, 7690 Bucharest, Romania\\
$^{e}$Fachbereich Physik, Universit\"at Siegen, 57068 Siegen, Germany\\
$^{f}$Fachbereich Physik, Universit\"at Wuppertal, 42097 Wuppertal, Germany\\
$^{g}$Istituto di Fisica dello Spazio Interplanetario, INAF, 10133 Torino, Italy\\
$^{h}$Soltan Institute for Nuclear Studies, 90950 Lodz, Poland\\
$^{i}$Department of Physics, University Bucharest, 76900 Bucharest, Romania}

\begin{abstract}
KASCADE-Grande is a large detector array dedicated for studies of
high-energy cosmic rays in the primary energy range from 100 TeV
to 1 EeV. The multi-detector concept of the experimental set-up
offers the possibility to measure simultaneously various observables
related to the electromagnetic, muonic, and hadronic air shower
components.
The experimental data are compared to predictions of CORSIKA simulations using
high-energy hadronic interaction models (e.g. QGSJET or EPOS), as well as
low-energy interaction models (e.g. FLUKA or GHEISHA).
This contribution will summarize the results of such investigations. In
particular, the validity of the new EPOS version 1.99 for EAS with energy
around 100 PeV will be discussed.
\end{abstract}

\maketitle

\thispagestyle{fancy}

\section{Introduction}
Studies of high-energy cosmic radiation by means of extensive air
shower (EAS) techniques require a proper understanding of high-energy
interactions in the Earth's atmosphere: Inferring the 
properties of the primary particles from EAS measurements one
relies on the simulations of the air shower development, whose
backbone is the hadronic cascade in the atmosphere.
Significant progress has been made 
during recent years to interpret air shower data
and main physical properties of the primary cosmic ray particles have been
measured.
In the energy range around 10$^{15}$\ eV,
energy spectra for elemental groups and mass
compositions of primary cosmic rays have been investigated.
Interpretations of these extensive air shower measurements are generally related to
air shower models to obtain physical properties of the shower including
primary particles.
Therefore, one of the goals of KASCADE-Grande is to investigate high-energy
interactions in the atmosphere and to improve contemporary models to describe
such processes.

The tests of hadronic interaction models require detailed measurements of
several shower observables.
The KASCADE-Grande experiment with its multi-detector concept of the experimental
set-up, measuring simultaneously the electromagnetic, muonic,
and the hadronic shower components, is particularly designed for such
investigations.

The KASCADE array measures an extensive air shower in the energy
range of 10$^{14}$ to 8$\times$10$^{16}$\ eV
and consists of 252 scintillator detector stations with
unshielded and shielded detectors located on a grid of 200$\times$200\ m$^{2}$
for the measurement of the electromagnetic and muonic shower components
independently.
In its center, an iron sampling calorimeter of 16$\times$20\ m$^{2}$ area
detects hadronic particles.

The KASCADE-Grande \cite{Apel}
array covering an area of 700$\times$700\ m$^{2}$ is
optimized to measure extensive air showers up to primary energies of 1\ EeV.
It comprises 37 scintillation detector stations located on a
hexagonal grid with an average spacing of 137\ m for the measurements of the
charged shower component. Each of the detector stations is equipped with
plastic scintillators covering a total area of 10\ m$^{2}$.

\section{Monte-Carlo simulations} 
The principal idea of the tests of hadronic interaction models is to simulate
air showers initiated by protons and irons nuclei as the two extremes of
possible primary particles
and to compare those simulated showers with the measurements.
For the air shower simulations the program CORSIKA \cite{Heck} has been used,
applying different embedded hadronic interaction models.
To determine the signals in the individual detectors, all secondary
particles at the ground level are passed through a detector simulation program
using GEANT package.
The predicted observables at ground level, such as e.g. the number of
electrons, muons and hadrons 
are then compared to the measurements.

FLUKA \cite{Fasso} and GHEISHA \cite{Fesefeldt} 
(E $<$ 200\ GeV and 80\ GeV, respectively)
models have been used for hadronic interactions at low energies.
High-energy interactions were treated with different models QGSJET-II-2
\cite{Ostapchenko} and EPOS 1.99 \cite{Pierog}.
Showers initiated by primary protons and iron nuclei have been simulated. The
simulations covered the energy range of 10$^{14}$ - 10$^{18}$ eV with zenith
angles in the interval 0$^{\circ}$ - 42$^{\circ}$. The spectral index in the
simulations was -2 and for the analysis it is converted to a slope of -3. 
The simulated events are
analyzed by the same method as the experimental data, in order to
avoid biases by pattern recognition and reconstruction algorithms.

\section{Investigations with KASCADE}
Using KASCADE measurements, the hadronic interaction models
QGSJET \cite{Kalmykov} (versions 98 and 01),
SIBYLL \cite{Engel, Ahn} (versions 1.6 and 2.1),
DPMJET \cite{Ranft}, VENUS \cite{Werner} and NEXUS \cite{Drescher} 
have been investigated.
First tests \cite{Antoni} supported QGSJET 98 as being most
compatible with the KASCADE data. Similar conclusions have been later
drawn for QGSJET 01 \cite{Apel2}. For the next model version, QGSJET-II-2
\cite{Ostapchenko}, some problems with the electron-hadron correlations have
been observed \cite{Hoerandel}.
Predictions of SIBYLL 1.6 were not compatible with air shower data, in
particular there were strong inconsistencies for hadron-muon correlations. 
These observations improved the development of SIBYLL 2.1 \cite{Ahn}.
The predictions of this model are very successful and
fully compatible with KASCADE air shower data \cite{Apel2}.
Investigations of the DPMJET version 2.5 
yield significant problems particularly for hadron-muon correlations, while the
newer version DPMJET 2.55 is found to be compatible with air shower data
\cite{Apel2}.
The predictions of the VENUS model revealed some inconsistencies in
hadron-electron correlations \cite{Antoni}.
The predictions of NEXUS 2 were found to be inconsistent with the KASCADE
data in particular for the investigations of hadron-electron correlations \cite{Apel2}.
Presently, the most compatible predictions are obtained from the models QGSJET
01 and SIBYLL 2.1.

Predictions of the interaction model EPOS 1.61 have been recently compared to
KASCADE air shower data \cite{Apel3}.
This model is a recent development, historically emerging from the VENUS and
NEXUS models.
This analysis indicates that the EPOS 1.61 delivers not enough hadronic energy
to the observation level, and also the energy per hadron seems to be too
small.
Presumably, the inconsistency of the EPOS predictions with the KASCADE
measurements is caused by too high inelastic cross sections for hadronic
interactions implemented in the EPOS model.
To solve these problems,
the treatment of screening effects in nuclear collisions has been improved in
EPOS. The new version EPOS 1.99 \cite{Pierog} has a reduced cross section and
inelasticity, compared to the previous EPOS 1.61 which leads to deeper shower
development.

\begin{figure}
\includegraphics[width=80mm]{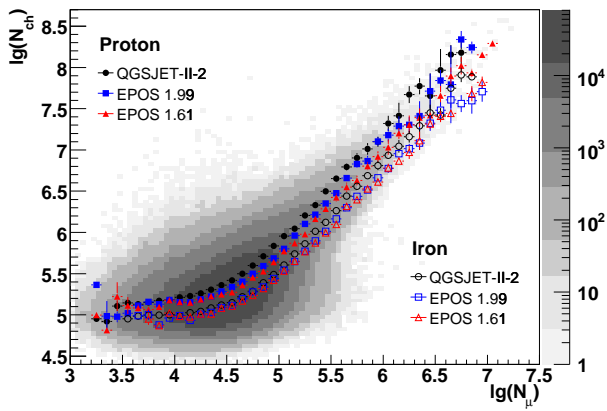}
\caption{Two-dimensional shower size spectrum measured by KASCADE-Grande,
  together with proton and iron induced showers for QGSJET-II-2 and EPOS
  simulations.}
\label{fig1}
\end{figure}

\section{Hadronic models QGSJET-II-2 and EPOS 1.99}
QGSJET is essentially based on the Quark-Gluon-String model
approach to high-energy hadronic interactions, including
a generalization of the latter for nucleus-nucleus collisions
and a treatment of semihard processes, using the so-called
semihard Pomeron approach \cite{Kalmykov}. The new hadronic interaction model
QGSJET-II accounts for non-linear interaction effects, which allows
one to obtain a consistent description of hadronic cross sections
and parton distribution functions \cite{Ostapchenko}.
EPOS \cite{Pierog} is a consistent quantum mechanical multiple scattering approach based on
partons and strings, where cross sections and the particle production are
calculated consistently, taking into account energy conservation in both cases.

Predictions of QGSJET-II-2 and EPOS 1.99 models have been investigated with
KASCADE-Grande data.
Figure \ref{fig1} represents the measured two-dimensional shower size spectrum, color
coded area, which includes full detector response by simulations.
The symbols correspond to 
the primary protons and iron
nuclei, as predicted by the interaction models EPOS 1.99 and QGSJET-II-2. 
The errors of mean values are plotted in Fig.\ \ref{fig1}.
It is shown that the most probable values 
for EPOS are shifted toward the higher muon numbers
with respect to QGSJET. 
This behavior implies, if EPOS predictions are used to derive the mass of primary
particles from the observed data, a dominantly light mass composition.
Air showers simulated with EPOS 1.99 have about 10\% more charged particles and
about 15\% less muons than QGSJET-II-2 at KASCADE-Grande energies.

\begin{figure}
\includegraphics[width=77mm]{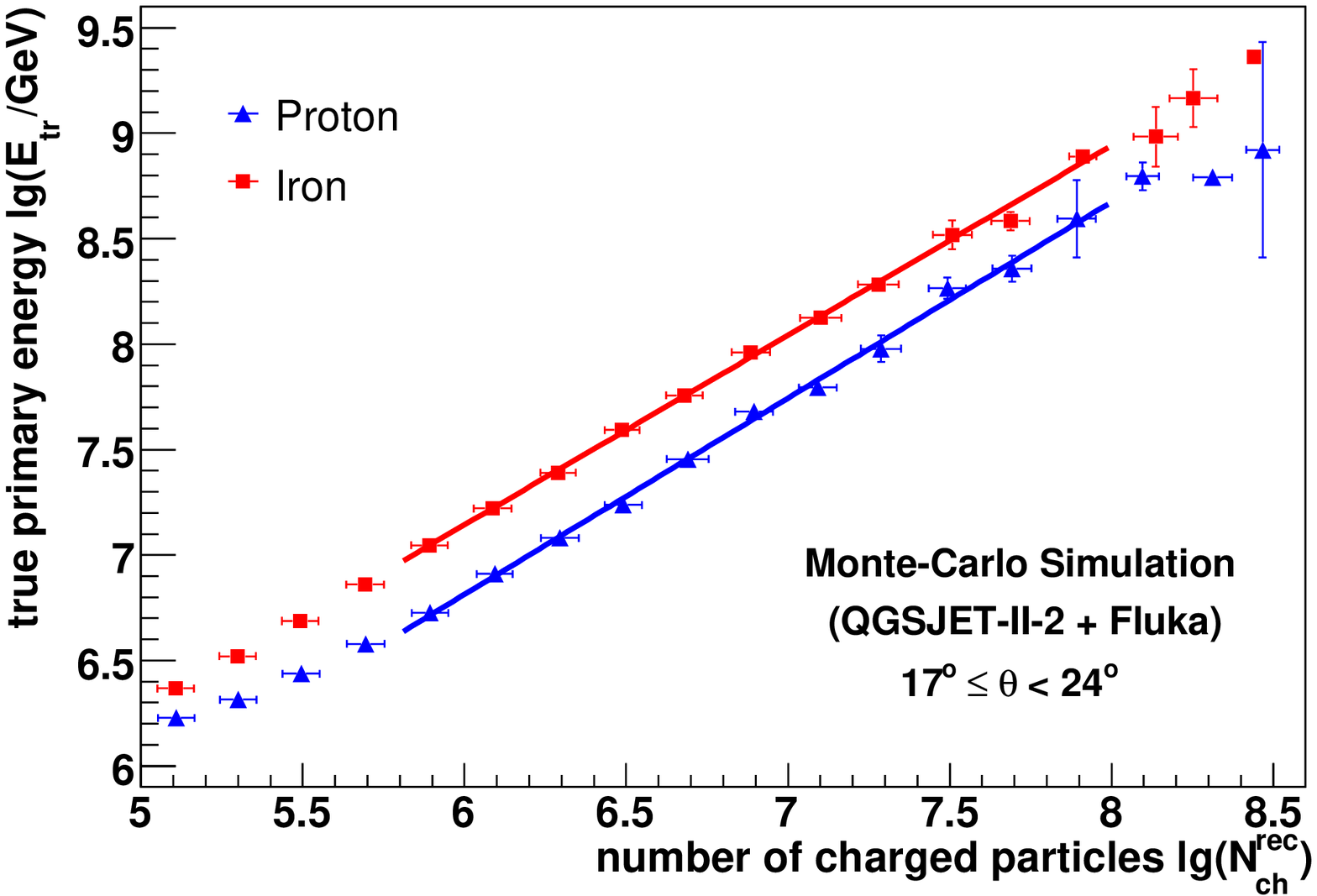}
\includegraphics[width=77mm]{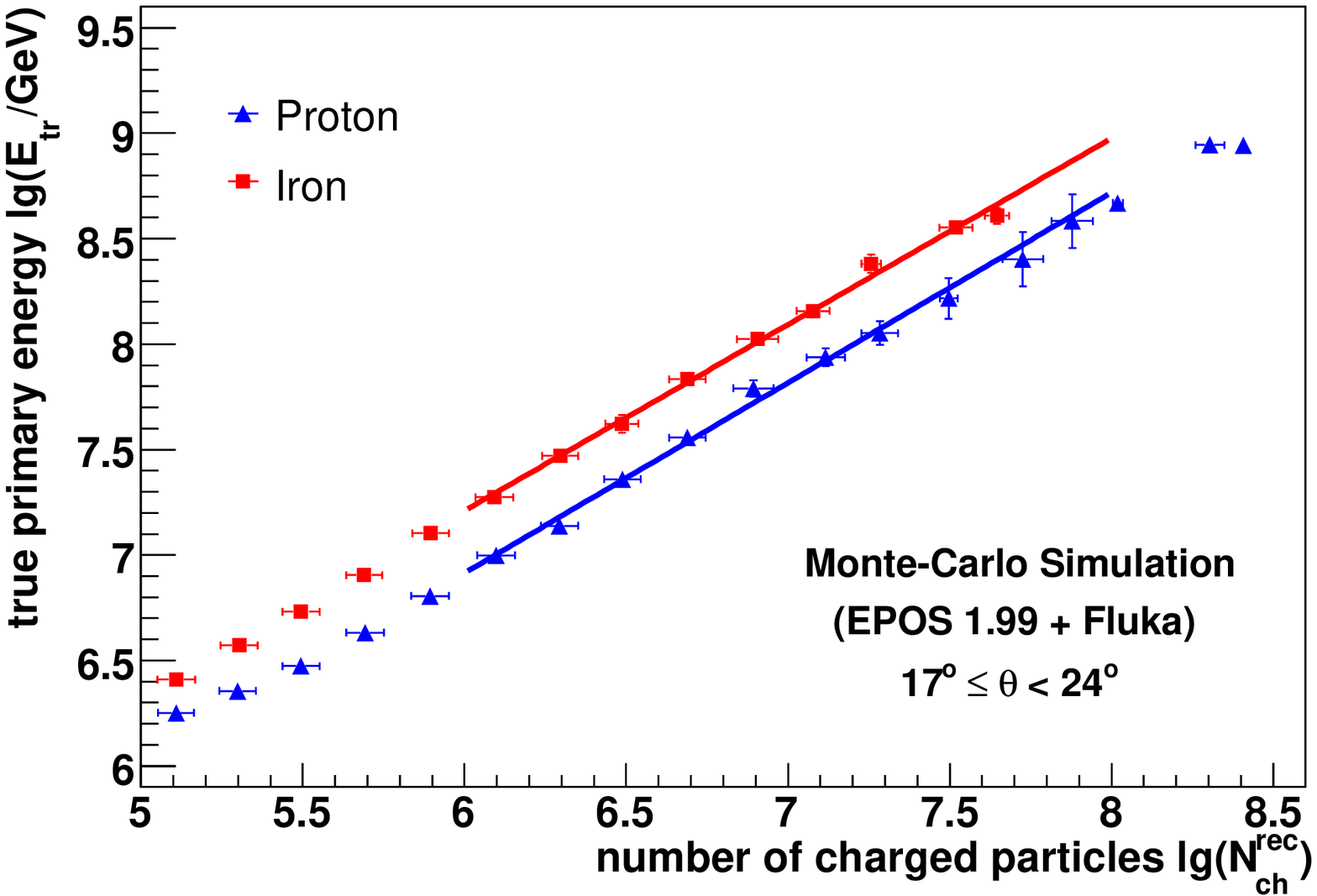}
\caption{The primary energy as a function of the number of charged particles
  for assumed pure proton and iron components for QGSJET-II-2 (top panel) and EPOS
  1.99 (bottom panel), respectively. The lines show the applied fits to the points.}
\label{fig2}
\end{figure}

The influences of the different hadronic interaction models on
the reconstructed all-particle energy spectrum was investigated
by performing the reconstructed
charged particle shower size method, based on simulations with the hadronic
interaction models QGSJET-II-2 and EPOS 1.99.
The shower size per individual event is corrected for attenuations in the
atmosphere by the constant intensity cut method and calibrated by Monte-Carlo
simulations under the assumption of a dependence $lgE \propto lgN_{ch}$ 
and a particular primary composition.
To determine the energy conversion relation between the number of charged
particles and the primary energy, the simulations were used.
The relation of the primary energy as a function of the number of charged
particles is shown in Fig.\ \ref{fig2} for the assumption of primary protons and iron,
respectively.
Assuming a linear dependence in logarithmic scale of $lgE = a + b \cdot
lgN_{ch}$, the fit is applied in the range of full trigger and reconstruction
efficiencies.
A detailed process for the energy reconstruction is given in Ref.\ \cite{Kang}.

\begin{figure}
\includegraphics[width=77mm]{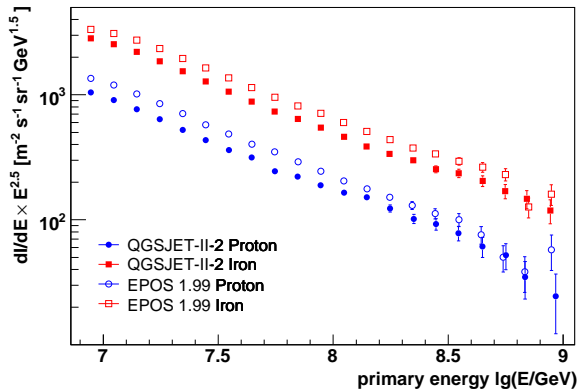}
\caption{Reconstructed all-particle energy spectra from KASCADE-Grande shower
  size for assuming proton and iron composition, based on two different
  hadronic interaction models QGSJET-II-2 and EPOS 1.99.}
\label{fig3}
\end{figure}

Figure \ref{fig3} shows the all-particle energy spectra obtained after applying the
energy reconstruction
functions as well as the appropriate correction for the bin to bin
fluctuations,
based on the assumption of iron and proton for QGSJET-II-2 and EPOS models. 
EPOS results lead to significantly higher flux (10-15\%) compared to QGSJET.
Since the EPOS model has 
for a fixed primary energy less charged particles,
it assigns higher flux.
As the calibration depends on simulations, other interaction models would
change the interpretation of KASCADE-Grande data,
so that more investigations by applying and comparing various
interaction models are needed.
 
\section{Muon density investigations}
\begin{figure}
\includegraphics[width=85mm]{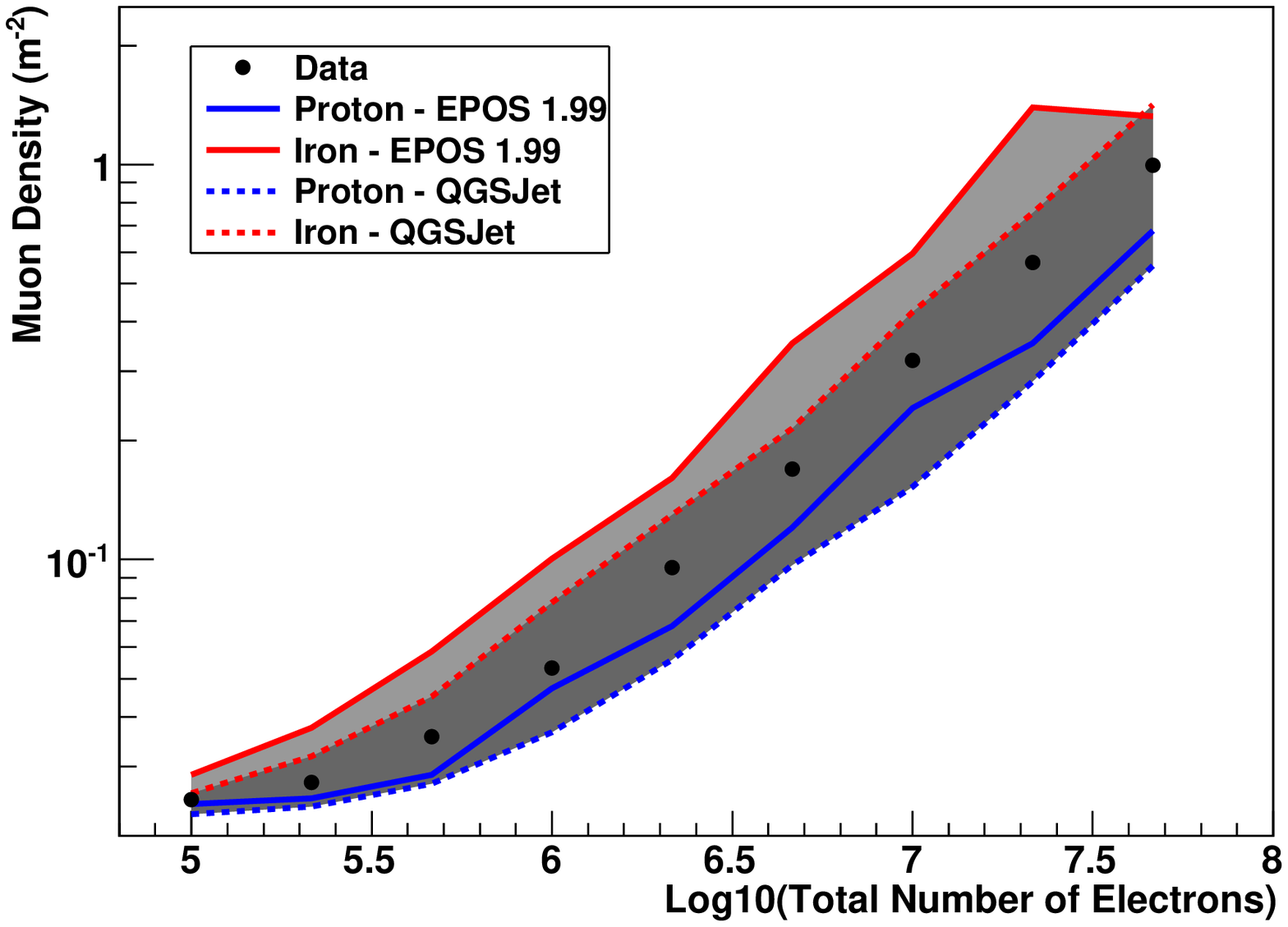}
\includegraphics[width=85mm]{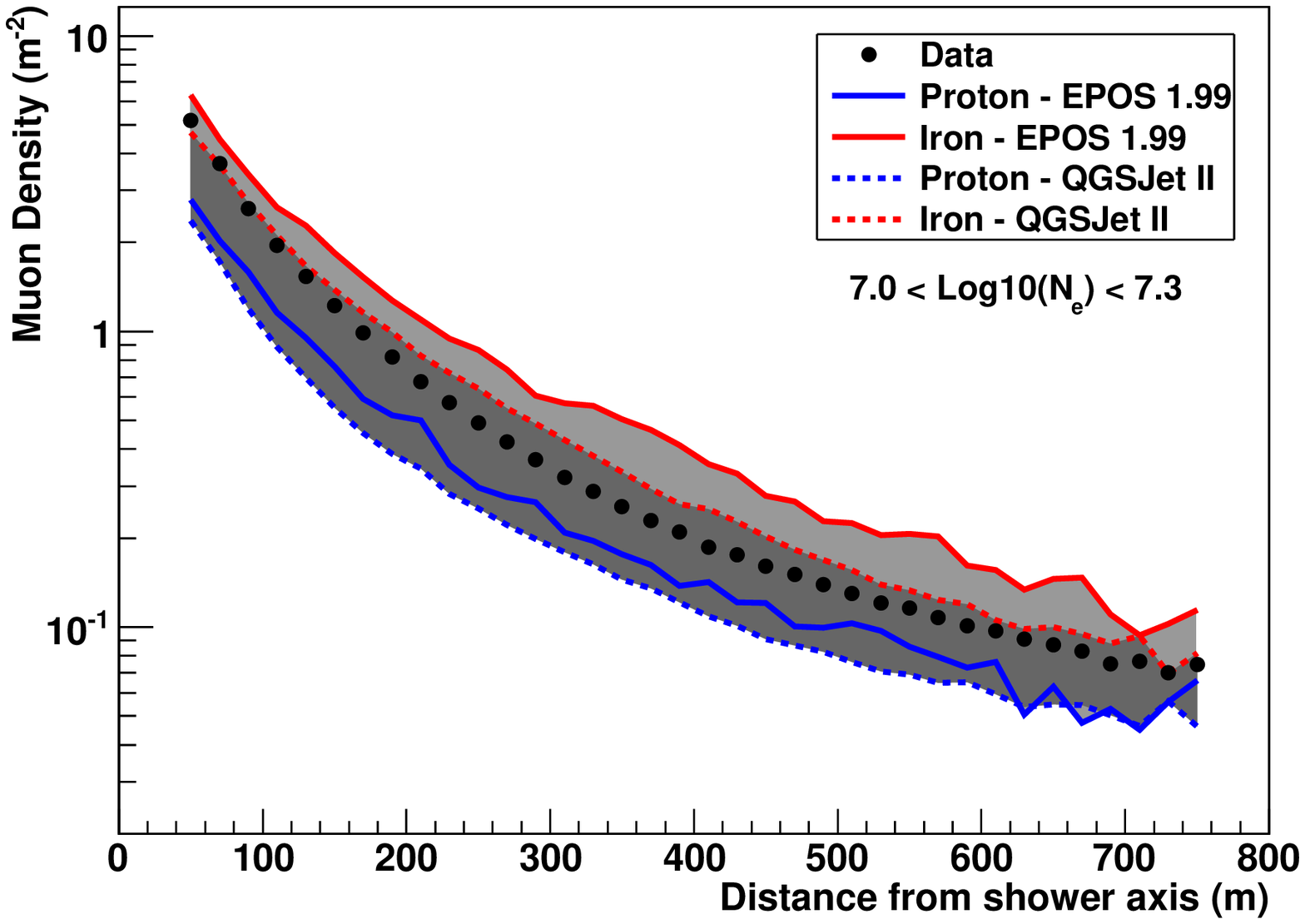}
\caption{Muon density as a function of the total number of electrons (top panel) and
  lateral distribution of muons (bottom panel) compared to the predictions of
  QGSJET-II-2 and EPOS 1.99.}
\label{fig4}
\end{figure}

The muon density can be directly measured by KASCADE, so that the
composition studies as well as the tests of hadronic interaction models with
muon densities can be performed \cite{Souza}.
Figure \ref{fig4} shows the correlation of muon density with 
the electron numbers and the distance from the shower axis,
compared to the predictions of
QGSJET-II-2 and EPOS 1.99 using proton and iron nuclei as primary particles.
The muon density decreases with increasing distance from the shower axis and
increases with increasing electron number.
Since an equal probability trigger for protons and iron primaries as a
function of distance from the shower axis is assumed, one should expect the lateral
density distribution to be parallel to pure composition primaries.
It shows that the lateral distributions of simulated proton and iron
shower are parallel, while the measured one is not quite parallel to the both
QGSJET-II-2 and EPOS 1.99 curves.
However, the lateral muon density distribution has a better slope than the
other models such as EPOS 1.61.
The QGSJET-II-2 model could fit the data with an intermediate primary
abundance between proton and iron nuclei, whereas
EPOS 1.99 would require abundance of light primary particles in order to fit
the data.

\section{Conclusion}
Testing of hadronic interaction models
QGSJET-II-2 and EPOS 1.99 implemented in the CORSIKA program
have been performed with KASCADE-Grande air shower data in the energy range of
10$^{16}$ to 10$^{18}$ eV.
From the muon density investigations,
the EPOS 1.99 model indicates that light abundances of primary cosmic ray
particles would be needed to fit the data.
On the other hand, the QGSJET-II-2 model describes the data with an
intermediate primary abundance between proton and iron nuclei.
The reconstructed all-particle energy spectra 
are presented by using the hadronic interaction models QGSJET-II-2 and EPOS 1.99.
The resulting spectra show that
the interpretation of the KASCADE-Grande data with EPOS 1.99 leads to
significantly higher flux as compared to the QGSJET-II-2 result.
Mored detailed investigations of EPOS 1.99 is still in work.

\end{document}